\documentclass[aps,preprint,nofootinbib]{revtex4}
\usepackage{amsmath,amssymb,mathtools,bm}
\usepackage{graphicx}
\usepackage{booktabs}
\usepackage{array}
\usepackage{xcolor}
\usepackage{hyperref}
\usepackage{placeins}
\usepackage{flafter}

\allowdisplaybreaks[4]

\newcommand{\dd}{\mathrm d}
\newcommand{\Mcal}{\mathcal M}
\newcommand{\xc}{x_{\gamma,\min}}
\newcommand{\xgc}{x_{g,\min}}
\newcommand{\BR}{\mathcal B}

\begin{document}

\title{True Leptonium ($l^+ l^-$) Production in UPC Triphoton Interaction\\[0.7cm]}

\author{\vspace{0.3cm} Qi-Ming Feng$^{1}$, Qi-Wei Hu$^{1,2}$, and Cong-Feng Qiao$^{1,3}$\footnote[1]{qiaocf@ucas.ac.cn}\vspace{0.3cm}}
\affiliation{\small {$^1$ School of Physical Sciences, University of Chinese Academy of Sciences, Beijing 100049, China\\
$^2$ Institute of High Energy Physics, Chinese Academy of Sciences, Yuquan Road 19B, Beijing 100049, China\\
$^3$ International Centre for Theoretical Physics Asia-Pacific, University of Chinese Academy of Sciences, Beijing 100190, China}}

\author{~\\}

\begin{abstract}
True leptonium states ($l^+ l^-$) are compact pure QED systems, first theoretically predicted eight decades ago. Although considerable efforts have been devoted to their search, only positronium has been experimentally confirmed shortly after its theoretical prediction. By contrast, dimuonium ($\mu^+ \mu^-$) and tauonium ($\tau^+ \tau^-$) remain unobserved to date, partly due to their low production yields. In this work, we investigate ortho-leptonium production through triphoton interactions in ultraperipheral heavy-ion collisions (UPCs). Two photons originate from different ions in one beam, while the third is supplied by an ion in the opposing beam. Within the equivalent photon approximation, we compare continuum dimuon and $J/\psi$ production with UPC measurements and evaluate ortho-leptonium production at the LHC and FCC. The results illustrate the possible role of triphoton fusion in leptonium production and the sensitivity of the calculated rates to the low-energy photon input.
\end{abstract}
\maketitle
\clearpage

\section{Introduction}
\label{sec:intro}
In high-energy collider experiments, scattering processes are overwhelmingly described in terms of two-particle collisions. Reactions involving three or more particles in the initial state are usually strongly suppressed, as the probability for multiple particles to interact within the same spacetime region is small. Nevertheless, multi-particle reactions can become phenomenologically relevant in the presence of strong enhancement mechanisms. Examples include the triple-$\alpha$ process in nuclear astrophysics~\cite{Nguyen:2011aa} and the $3\to2$ reactions appearing in certain dark matter scenarios~\cite{Hochberg:2014dra}.

A similar possibility arises in collider environments through ultraperipheral collisions (UPCs) of relativistic heavy ions. In UPCs, the impact parameter exceeds the sum of the nuclear radii, suppressing hadronic interactions while preserving the long-range electromagnetic fields of the ions. Owing to their large charge, heavy ions generate intense electromagnetic fields, which can be interpreted as fluxes of equivalent photons with strengths scaling approximately as $Z^2$~\cite{Budnev:1975poe}. Consequently, heavy-ion UPCs provide large photon luminosities and a favorable environment for studying photon-induced reactions at high energies.

Over the past decades, photon-induced reactions in UPCs have been extensively studied at RHIC and the LHC~\cite{Bertulani:2005ru,Baltz:2007kq,STAR:2004bzo,ATLAS:2017fur,CMS:2018erd,ALICE:2012yye,ALICE:2015nbw,CMS:2016itn}. Many of these processes are described by the equivalent photon approximation (EPA), originally formulated by Weizs\"acker and Williams, which treats the electromagnetic fields of relativistic ions as fluxes of quasi-real photons~\cite{Baltz:2007kq,Klein:1999qj,vonWeizsacker:1934nji,Williams:1934ad}.

The large photon fluxes in heavy-ion UPCs naturally raise the possibility of multiphoton fusion processes. Three-photon production of ortho-positronium in two-ion collisions was studied by Kotkin et al.~\cite{Kotkin:1998hu}, including Coulomb corrections. Bertulani and Navarra~\cite{Bertulani:2001zk} examined two- and three-photon meson production using the nuclear fields, and subsequent work has considered related multiphoton channels~\cite{Francener:2021wzx,Fariello:2023uvh}. In the leading three-photon contributions of these calculations, two photons couple to the same nucleus. Here, by contrast, each photon is supplied by a different ion: two ions belong to one bunch and the third to the opposing bunch. The additional ion contributes through its electromagnetic field in the production region. All three inputs enter one connected amplitude, unlike double-parton scattering (DPS)~\cite{Diehl:2011tt,Blok:2013bpa}.

A particularly appealing application of three-photon fusion is the production of leptonium bound states, i.e., electromagnetic bound states of a lepton--antilepton pair. While positronium ($e^+e^-$) has long served as a precision laboratory for QED~\cite{Pirenne1946,Deutsch:1951zza}, heavier analogues such as dimuonium ($\mu^+\mu^-$) and tauonium ($\tau^+\tau^-$) remain experimentally elusive. In recent years, their production has attracted growing theoretical attention~\cite{Bilenky:1969zd,Hughes1971,Malenfant:1987tm,Jentschura:1998vkm,Brodsky:2009gx,Mohsen2015,Lamm:2016djr,dEnterria:2022ysg,dEnterria:2023yao,Shao:2022cly,Bertulani:2023nch,Dai:2024imb,Francener:2024eep,Martynenko:2024rfj,dEnterria:2025ecx,Gargiulo:2025pmu,Zhao:2025gia,Feng:2025ppc,ColpaniSerri:2025vdz}. The large photon luminosities in UPCs motivate an examination of their production through multiphoton fusion.

The ground states of leptonium are classified into spin-singlet (para) and spin-triplet (ortho) configurations. For dimuonium, these correspond to the ${}^1S_0$ and ${}^3S_1$ states. Although para-dimuonium production in UPCs has been shown to yield sizable rates, its dominant diphoton decay is experimentally challenging to reconstruct because of the softness of the final-state photons~\cite{dEnterria:2025ecx}. By contrast, ortho-dimuonium predominantly decays into an $e^+e^-$ pair, providing a complementary experimental signature. The triphoton channel considered here complements conventional radiative production~\cite{Feng:2025ppc}. Its observable yield also depends on decay acceptance and the dissociation of ortho-dimuonium in detector material~\cite{CidVidal:2019qub}.

In this work, we investigate leptonium production in ultraperipheral heavy-ion collisions from the perspective of three-particle initial states within the framework of nonrelativistic QED (NRQED)~\cite{Caswell:1985ui}. We formulate the three-particle contribution and compare continuum dimuon and $J/\psi$ production with UPC data after adding the conventional contributions. We then apply the same source description to ortho-leptonium and study the dependence of the production rates on the low-energy photon input.

\section{Triphoton interaction mechanism}
\label{sec:mechanism}
In heavy-ion UPCs, three-photon fusion is a leading Born-level electromagnetic channel for producing an ortho-vector state. Direct two-photon production of such a state is forbidden by charge-conjugation symmetry and the Landau--Yang theorem~\cite{Landau:1948kw,Yang:1950rg}. A conventional electromagnetic channel is the radiative $2\to2$ process, $\gamma\gamma\to V+\gamma$. The product of three coherent single-ion photon fluxes scales as $Z^6$, enhancing the three-photon channel for heavy nuclei. The direct $3\to1$ process, $3\gamma\to V$, also has no final-state recoil photon, in contrast to the radiative channel.

We describe the three-particle interaction with wave-packet initial states in the $S$-matrix formalism. Taking particle 3 as the transverse reference, we specify the incoming configuration by the impact parameters of particles 1 and 2 and define the effective reaction quantity $\Sigma_3$ by integrating the transition probability over both impact parameters.

Thus,
\begin{equation}
\Sigma_3=\int\dd^2b_1\dd^2b_2\,\mathcal P(\bm b_1,\bm b_2),
\label{eq:sigma_def}
\end{equation}
where $\bm b_1$ and $\bm b_2$ denote the impact parameters of particles 1 and 2 relative to particle 3. Here, $\mathcal P(\bm b_1,\bm b_2)$ denotes the transition probability integrated over the final-state momenta.

To implement this picture at the amplitude level, we construct the initial state as a superposition of wave packets and evaluate the corresponding $S$-matrix element. After integration over the transverse impact parameters, the transverse momenta of particles 1 and 2 are equal in the amplitude and its complex conjugate.

Carrying out the remaining integrations (see Appendix~\ref{app:wavepackets} for details), the differential effective cross section can be expressed as
\begin{equation}
\dd\Sigma_3=\frac{|\Mcal|^2}{4E_1E_2|v_1-v_2|}
\frac{\dd k_{3z}}{(2\pi)2E_3}\,\dd\Phi_n,
\label{eq:dsigma}
\end{equation}
where $\dd\Phi_n$ denotes the $n$-body phase space of the final state, and $v_i=\partial E_i/\partial k_{iz}$ are the longitudinal velocities of the incoming particles. For the photon kinematics considered here, particles 1 and 2 move in opposite beam directions.

In collider kinematics, it is convenient to express the remaining longitudinal degree of freedom in terms of the momentum fraction of the corresponding beam,
\begin{equation}
k_{3z}=x_3P_z^{\rm beam},\qquad
\frac{\dd k_{3z}}{E_3}\to\frac{\dd x_3}{x_3},
\label{eq:x_def}
\end{equation}
which leads to the final form
\begin{equation}
\dd\Sigma_3=\frac{|\Mcal|^2}{16\pi E_1E_2|v_1-v_2|}
\frac{\dd x_3}{x_3}\,\dd\Phi_n.
\label{eq:dsigma_x}
\end{equation}

In UPCs, the incoming photons are generated by the electromagnetic fields of relativistic ions. Their flux is described by the equivalent photon approximation (EPA). Using independent single-ion spectra, the three-photon convolution is written as
\begin{equation}
\begin{aligned}
\dd\Sigma_3(\mathrm{ion}+\mathrm{ion}\to f)
&=\int\dd x_1\,\dd x_2\,\dd x_3\,
 \\[3pt]
&\quad\times f_\gamma(x_1)f_\gamma(x_2)f_\gamma(x_3)
 \frac{|\Mcal|^2}{8\pi x_3s_{12}}\,\dd\Phi_n,
\end{aligned}
\label{eq:epa_conv}
\end{equation}
where $s_{12}=(k_1+k_2)^2$. The photon distribution function is given by~\cite{Baur:2001jj}
\begin{equation}
f_\gamma(x)=\frac{2\alpha Z^2}{\pi x}
\left[\eta K_0(\eta)K_1(\eta)
-\frac{\eta^2}{2}\bigl(K_1^2(\eta)-K_0^2(\eta)\bigr)\right],
\label{eq:flux}
\end{equation}
with $\eta\equiv xm_pR$, where $m_p$ is the proton mass, $R$ is the ion radius, and $K_0,K_1$ are the modified Bessel functions. For Pb, the numerical calculation uses $Z=82$ and $\eta=33.76x$.

A crucial feature of the triphoton process is that the three nuclear centers need not coincide at the production point. The electromagnetic field of an additional ion can contribute in the region where the fields of the selected UPC pair overlap. In the present approximation, we describe this overlap through the effective source density entering the three-particle luminosity.

Within the adopted source-overlap model, the three-particle luminosity is
\begin{equation}
\mathcal L_3=\frac{N}{\Sigma_3}
=\frac{N_1N_2N_3}{\hat S^2}
=\frac{N_3}{\hat S}\mathcal L_2,
\label{eq:lumi3}
\end{equation}
where $N$ denotes the total number of events, $N_i$ is the source population assigned to input $i$, and $\hat S=4\pi\sigma_x\sigma_y$ is the effective transverse area for a Gaussian beam profile, with $\sigma_x$ and $\sigma_y$ the RMS transverse beam sizes. Here, $\mathcal L_2$ represents the conventional two-particle luminosity. Both beam assignments are included in $\mathcal L_3$.

For phenomenological applications, we define the effective three-particle cross section
\begin{equation}
\hat\sigma_3=\frac{\Sigma_3N_3}{\hat S}
=\frac{\Sigma_3N_3}{4\pi\sigma_x\sigma_y},
\label{eq:sighat}
\end{equation}
which satisfies $N=\mathcal L_2\hat\sigma_3$. Thus $\hat\sigma_3$ is a bunch-dependent effective cross section. We use $N_i=1.5\times10^8$ and $\sigma_x=\sigma_y=3\,\mu\mathrm m$ for both energy settings, keeping the additional source density fixed.

\section{The $3\to1$ processes}
\label{sec:production}
Applying the three-particle formalism with the leading nonrelativistic bound-state projection, the squared amplitude for producing a vector leptonium state from three photons in the collinear configuration is
\begin{equation}
|\Mcal|^2=\frac18\frac{128e^6}{m_\ell^3}|\Psi(0)|^2,
\label{eq:hard}
\end{equation}
where $|\Psi(0)|^2=\alpha^3m_\ell^3/(8\pi)$ is the squared leptonium wave function at the origin. For $J/\psi$ production, the squared amplitude acquires the color-charge factor $C_AQ_c^6$, where $C_A=3$ and $Q_c=2/3$. The lepton mass is replaced by the charm-quark mass, and the squared wave function by $|\Psi(0)|^2=9M_{J/\psi}^2\Gamma_{e^+e^-}/(64\pi\alpha^2)$.

For the $3\to1$ kinematics, we relabel the inputs of Eqs.~\eqref{eq:dsigma}--\eqref{eq:epa_conv} so that photons 1 and 2 co-propagate, while photon 3 moves in the opposite direction. Taking $x_1$ and $x_2$ as the independent momentum fractions, energy-momentum conservation fixes $x_3$ through the resonance condition
\begin{equation}
x_3=\frac{M^2}{(x_1+x_2)S},
\label{eq:resonance}
\end{equation}
where $M$ is the mass of the final-state vector particle and $S=s_{NN}$. By integrating over the respective photon distributions, the leptonium expression takes the form
\begin{equation}
\begin{aligned}
\dd\Sigma_3(\mathrm{ion}+\mathrm{ion}\to(\ell^+\ell^-)[{}^3S_1])
&=\int\dd x_1\,\dd x_2\,f_\gamma(x_1)f_\gamma(x_2)\\[3pt]
&\quad\times f_\gamma\!\left(\frac{4m_\ell^2}{(x_1+x_2)S}\right)
 \frac{8\pi^2\alpha^6}{m_\ell^2x_1x_2S}.
\end{aligned}
\label{eq:master}
\end{equation}
The cross section for $J/\psi$ production can be analogously obtained by substituting the corresponding mass and squared amplitude.

The EPA spectrum grows as $x\to0$. In two-photon production of a massive final state, the threshold constrains the product of the two momentum fractions. In the $3\to1$ process, however, either $x_1$ or $x_2$ can approach zero while their sum remains fixed. Fixing the final-state mass therefore does not remove the sensitivity to an individual soft photon.

Several physical scales enter the electromagnetic input. Nuclear coherence gives the characteristic upper photon-energy scale $\gamma/R$~\cite{Bertulani:2005ru,Baur:2001jj}. This is reflected in Eq.~\eqref{eq:flux}: the modified Bessel functions suppress the flux exponentially for $\eta>1$. The low-frequency part of the field, on the other hand, must be considered together with the reaction it induces. Electromagnetic dissociation and bound-free pair production illustrate how reaction thresholds restrict the contributing photon energies~\cite{ALICE:2022iqi,Aste:2007vs}. For the present process, the corresponding kinematic restrictions follow from the three-photon resonance condition.

Imposing a lower bound $x_i\geq\xc$ on each photon momentum fraction gives
\begin{equation}
2\xc<x_{12}<\frac{M^2}{\xc S},
\qquad \xc<\frac{M}{\sqrt{2S}},
\label{eq:domain}
\end{equation}
where $x_{12}=x_1+x_2$, with the individual upper bounds $x_i\leq1$ also understood. These conditions become particularly important for light bound states at high collision energies.

Including a final-state energy threshold $E_{\min}$ further modifies the allowed region. No additional constraint arises for $E_{\min}<M$, while for $E_{\min}>M$ the kinematic domain is restricted to
\begin{equation}
x_{12}<\frac{E_{\min}-\sqrt{E_{\min}^2-M^2}}{\sqrt S}
\quad\text{or}\quad
x_{12}>\frac{E_{\min}+\sqrt{E_{\min}^2-M^2}}{\sqrt S}.
\label{eq:energycut}
\end{equation}
The mass and energy conditions constrain $x_{12}$ and $x_3$, but leave the sharing between the two co-propagating photons undetermined. An individual lower bound therefore requires information beyond these kinematic relations.

Finite bunch dimensions, source-field overlap, and the produced wave packet can affect the low-energy contribution. Quantifying these effects requires source correlations that retain their frequency and momentum dependence~\cite{Wang:2021kxm,Kotkin:2003jz}. We examine the low-energy sensitivity by varying $\xc$, using the same photon range for the different final states.

\section{Phenomenology and discussion}
\label{sec:results}
To examine the phenomenological role of triphoton interactions and their dependence on the low-energy input, we consider continuum dimuon, $J/\psi$, and ortho-leptonium production in Pb--Pb UPCs. The experimental comparisons use LHC measurements at $\sqrt S=5.02\,\mathrm{TeV}$, while leptonium production is evaluated at the LHC and FCC ($\sqrt S=39.4\,\mathrm{TeV}$)~\cite{Schaumann:2015fsa}.

\begin{figure}[!b]
\centering
\includegraphics[width=\textwidth]{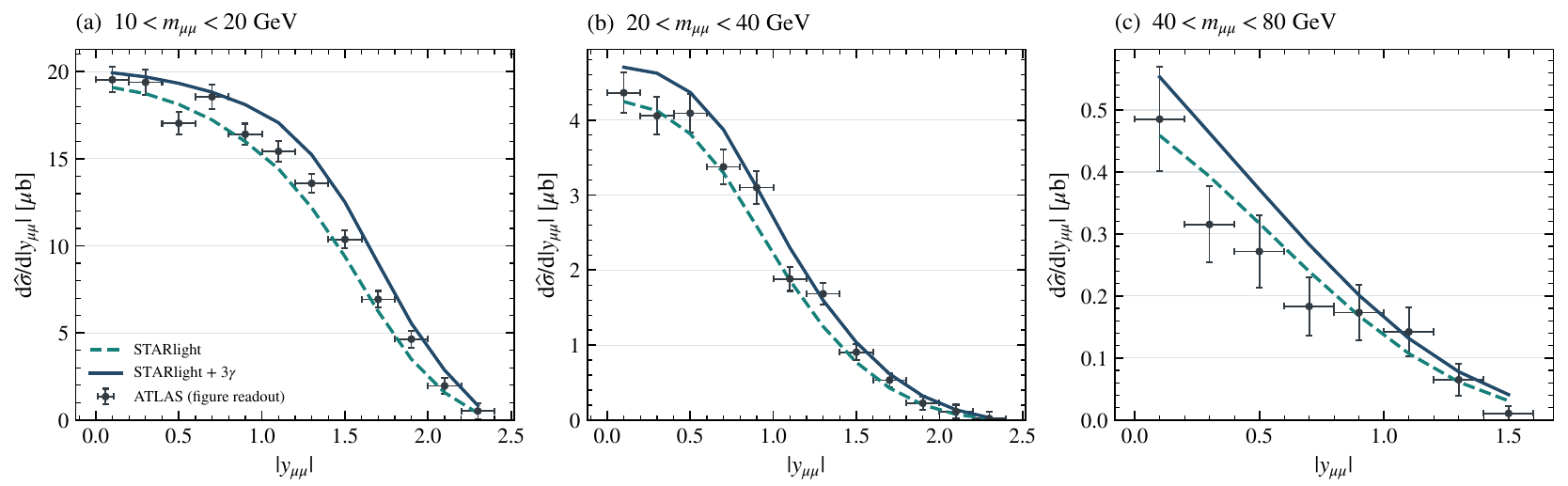}
\caption{Continuum-dimuon rapidity distributions in Pb--Pb UPCs at $\sqrt S=5.02\,\mathrm{TeV}$ with $x_{\gamma,\min}=10^{-6}$, compared with ATLAS data~\cite{ATLAS:2020epq}.}
\label{fig:continuum}
\end{figure}

\subsection{Continuum dimuon production}
Exclusive dimuon production in UPCs provides a well-measured comparison for the triphoton mechanism. The ATLAS Collaboration reported $\sigma^{\mu\mu}=34.1\pm0.3\,\text{(stat)}\pm0.7\,\text{(syst)}\,\mu\mathrm b$ in the fiducial region $p_{T,\mu}>4\,\mathrm{GeV}$, $|\eta_\mu|<2.4$, $m_{\mu\mu}>10\,\mathrm{GeV}$, and $p_{T,\mu\mu}<2\,\mathrm{GeV}$~\cite{ATLAS:2020epq}. Under the same conditions, STARlight predicts $32.1\,\mu\mathrm b$~\cite{Klein:2016yzr}, while gamma-UPC gives $36.2\pm4.2\,\mu\mathrm b$~\cite{Shao:2022cly}. The conventional two-photon mechanism thus accounts for most of the measured rate.

Using Eq.~\eqref{eq:epa_conv} in the collinear, massless-muon approximation with $\xc=10^{-6}$, we obtain a three-photon contribution of approximately $5.30\,\mu\mathrm b$. Adding it to STARlight gives $37.4\,\mu\mathrm b$, above the measured central value. Its mass and rapidity dependence is shown in Fig.~\ref{fig:continuum}.

The addition improves some bins but produces an excess in others across the three mass windows. Agreement in an integrated rate would therefore not ensure agreement in the differential distributions. Photon-flux geometry, hadronic survival, and radiative effects in the conventional calculation also enter the comparison~\cite{ATLAS:2020epq,Klein:2016yzr}.

\subsection{$J/\psi$ production}
In contrast to continuum dimuon production, coherent $J/\psi$ production is generally dominated by QCD-driven mechanisms, primarily photon--Pomeron exchange. Earlier calculations find a much smaller direct electromagnetic three-photon contribution~\cite{Bertulani:2001zk}. Our direct result is likewise small for the photon bound used below, so we compare the sum of conventional photoproduction and the additional channels with data.

Resolved photons provide gluon inputs to the hard process. For a color-singlet $J/\psi$ in the $3\to1$ channel without additional hard radiation, the single-resolved amplitude $\gamma\gamma g\to J/\psi$ vanishes. We include direct three-photon production together with double-resolved $\gamma gg\to J/\psi$ and triple-resolved $ggg\to J/\psi$ production. For each resolved leg, we independently convolve the nuclear photon flux with the pointlike component of the GRS leading-order gluon distribution of the photon~\cite{Gluck:1999ub}, evaluated at $Q^2=M_{J/\psi}^2$. We use $\alpha_s=0.2$ in the hard amplitudes. Double-resolved production dominates the additional contribution for the inputs considered.

\begin{figure}[htbp]
\centering
\includegraphics[width=0.98\textwidth]{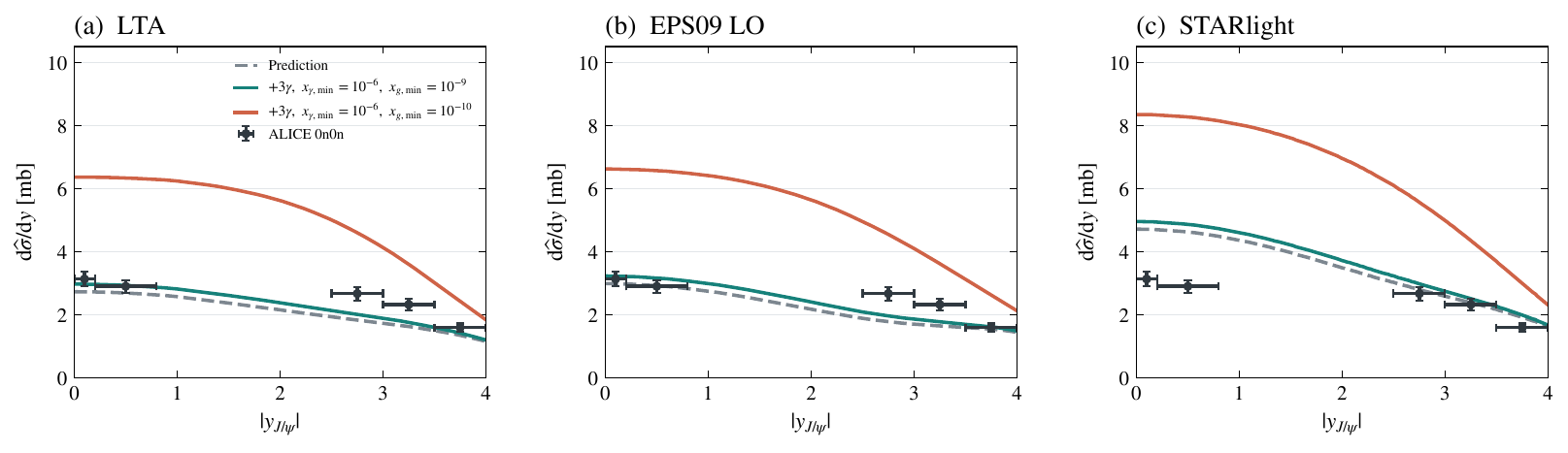}
\caption{LTA, EPS09 LO, and STARlight predictions including the full direct and resolved contributions, compared with ALICE $0n0n$ coherent $J/\psi$ data~\cite{ALICE:2023jgu}.}
\label{fig:jpsi}
\end{figure}

\begin{figure}[htbp]
\centering
\includegraphics[width=\textwidth]{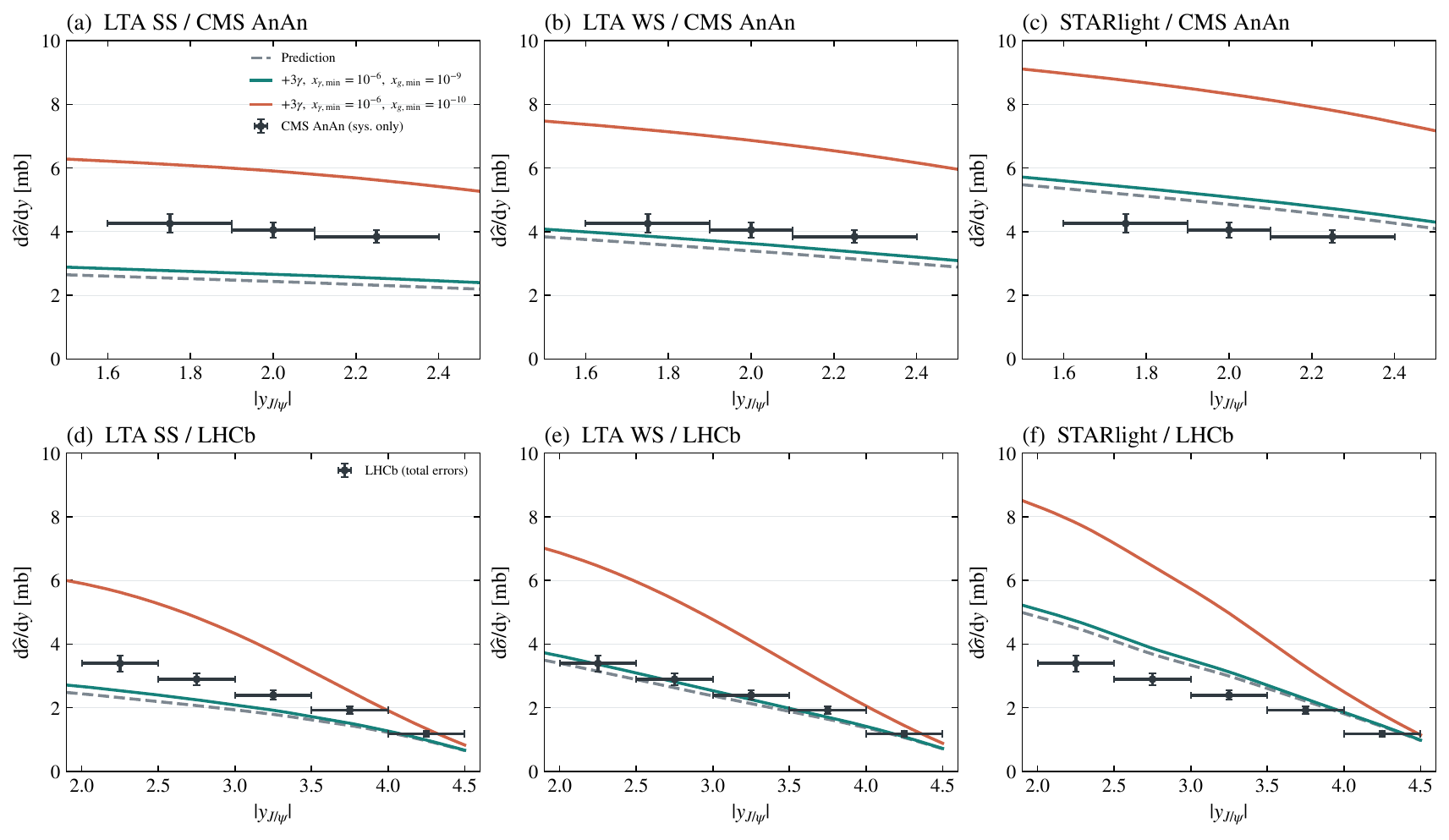}
\caption{LTA SS, LTA WS, and STARlight predictions including the full direct and resolved contributions, compared with CMS $AnAn$~\cite{CMS:2023snh} (upper panels) and LHCb~\cite{LHCb:2022ahs} (lower panels) coherent $J/\psi$ data.}
\label{fig:experiments}
\end{figure}

The gluon fraction $x_g$ is defined relative to its parent photon, so the beam-relative gluon fraction is $x_\gamma x_g$. In Figs.~\ref{fig:jpsi} and~\ref{fig:experiments}, we compare $\xgc=10^{-9}$ and $10^{-10}$ at $\xc=10^{-6}$. Both choices probe the small-$x_g$ extrapolation of the gluon parameterization~\cite{Gluck:1999ub}. For the ALICE comparison, the direct and resolved contributions are added separately to the leading-twist shadowing approach (LTA), EPS09 LO, and STARlight predictions in Ref.~\cite{ALICE:2023jgu}. In the figures, $+3\gamma$ denotes the full direct, double-resolved, and triple-resolved addition.

For the LTA baseline, $\xgc=10^{-9}$ improves the central region but leaves deficits in the forward bins. For EPS09 LO, the central baseline is already close to the data, so the same addition does not improve that region uniformly, although it reduces part of the forward deficit. Lowering the bound to $10^{-10}$ raises both predictions above the measured central and forward values. In the small-$x_g$ range examined, varying the bound mainly changes the magnitude of the additional term rather than its rapidity shape, leaving limited freedom to alter the relative central and forward yields. For the displayed STARlight baseline, an additional positive contribution increases the existing excess.

The comparison is extended to the CMS neutron-inclusive ($AnAn$) sample~\cite{CMS:2023snh} and the LHCb forward measurement~\cite{LHCb:2022ahs} at the same energy in Fig.~\ref{fig:experiments}. The conventional baselines are LTA with strong (SS) or weak (WS) shadowing and STARlight, taken from the CMS $AnAn$ analysis and used for both data sets. These differ from the $0n0n$ baselines in Fig.~\ref{fig:jpsi}; the added term uses the same photon and gluon inputs in the respective rapidity windows.

For CMS, the addition at $\xgc=10^{-9}$ reduces the deficits of both LTA predictions, which nevertheless remain below the three measured central values. The weak-shadowing baseline leaves less room for an additional term than the strong-shadowing baseline. At $\xgc=10^{-10}$, both predictions exceed the CMS central values.

The LHCb comparison shows a similar sensitivity farther forward. With weak shadowing, the smaller addition brings the lower-rapidity bins closer to the data, but its effect decreases toward the most forward region. The larger addition produces an excess, particularly at smaller rapidities, for both shadowing choices. The STARlight bin averages already exceed the CMS and LHCb central values, so a positive term increases that excess.

The addition reduces some differences between the data and the conventional predictions, but remains sensitive to the small-$x_g$ extrapolation. These comparisons do not determine a physically motivated common cutoff. The full additional contribution is assigned to each sample. Relating it to the measured yields requires nuclear excitation and photon remnants to be treated consistently with the respective event selections: ALICE selects $0n0n$, CMS includes all neutron-emission classes in $AnAn$, and LHCb applies an activity veto.

\subsection{Ortho-leptonium production}
\begin{figure}[htbp]
\centering
\includegraphics[width=0.96\textwidth]{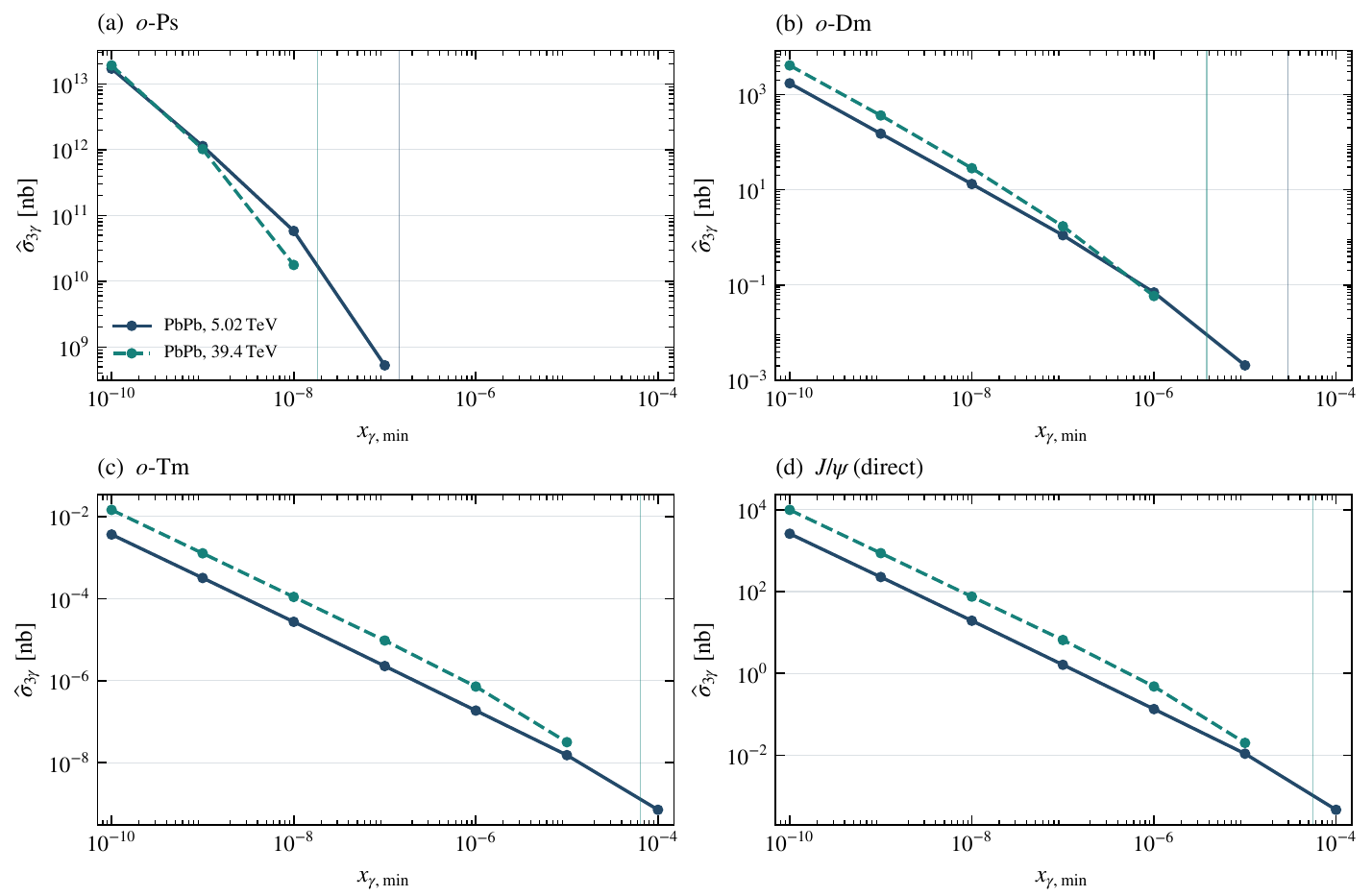}
\caption{Total cross sections for direct triphoton production of ortho-positronium, ortho-dimuonium, ortho-tauonium, and $J/\psi$ in Pb--Pb UPCs at the LHC and FCC. The color-matched vertical lines indicate the kinematic upper bounds $M/\sqrt{2S}$ for the respective collision energies.}
\label{fig:total}
\end{figure}

Using the same source description, we evaluate direct ortho-positronium (o-Ps), ortho-dimuonium (o-Dm), and ortho-tauonium (o-Tm) production over a range of $\xc$. The cross sections are shown in Fig.~\ref{fig:total}, together with the direct $J/\psi$ result. A strong mass hierarchy is apparent: positronium has the largest rate when its phase space is open, followed by dimuonium and tauonium. Both the mass dependence of the production kernel and the photon fractions required by Eq.~\eqref{eq:resonance} contribute to this hierarchy.

The different kinematic boundaries are especially significant for positronium. Owing to its small mass, the condition in Eq.~\eqref{eq:domain} closes the allowed region at a much lower photon-fraction bound than for the heavier states. This boundary moves to smaller $\xc$ as the collision energy increases. Thus the same photon range can allow dimuonium and tauonium production while leaving no positronium phase space. The suppression near these boundaries is kinematic and occurs in addition to the smooth decrease of the photon flux.

The higher collision energy does not enhance all channels uniformly. As follows from Eq.~\eqref{eq:domain}, the allowed range of $\xc$ becomes more restrictive as $\sqrt S$ increases. In particular, the FCC dimuonium curve is not everywhere above the LHC curve. Tauonium and direct $J/\psi$ production are enhanced at smaller $\xc$, while their FCC phase space closes earlier as $\xc$ increases. These trends arise from the interplay between the photon spectra and the available phase space.

For comparison, the two-ion calculation of Kotkin et al.~\cite{Kotkin:1998hu} gives an ortho-positronium cross section of $24\,\mathrm{mb}$ at $\gamma=3000$, including excited $S$ states and Coulomb corrections. The present result is for the ground state and additionally depends on the source density in Eq.~\eqref{eq:sighat} and the specified low-energy range.

At the lowest photon-fraction bounds, the large positronium rates require a treatment of multiple production. Unitarity constrains the probability at fixed source geometry, whereas the mean multiplicity may exceed unity~\cite{Alscher:1996mja}. The largest single-production estimates therefore cannot be interpreted as exclusive event rates without this treatment.

The low-energy dependence motivates a more differential description of the initial electromagnetic fields. Nuclear wave-packet and beam-correlation methods retain source-position and momentum correlations~\cite{Wang:2021kxm,Kotkin:2003jz}. Effective descriptions of extended charges~\cite{Plestid:2024ygd} and asymptotic soft interactions~\cite{Hannesdottir:2019opa} may also inform such a treatment, but their matching to the present three-source process remains to be established.

For dimuonium, including softer photons raises the calculated rate from the sub-nanobarn to the nanobarn and microbarn ranges. Its relative importance compared with radiative production~\cite{Feng:2025ppc} therefore depends on the low-energy input as well as the coherent charge enhancement. The predominantly electronic decay of ortho-dimuonium, with $\BR((\mu^+\mu^-)\to e^+e^-)\simeq0.86$~\cite{Karshenboim:1998am}, gives a useful experimental channel. Reconstruction of the soft pair is challenging, while the characteristic lifetime of order $10^{-12}\,\mathrm s$ provides a possible displaced-vertex signature~\cite{CidVidal:2019qub}. The observable yield depends on the decay acceptance, reconstruction efficiency, and survival in detector material.

\section{Summary and conclusions}
\label{sec:conclusion}
In summary, we have investigated vector-onium production through triphoton fusion in ultraperipheral heavy-ion collisions. The three photons are supplied by different ions in two colliding bunches. We use a wave-packet description and an effective source density to evaluate the corresponding production rates.

The continuum-dimuon comparison shows that the additional contribution improves some mass and rapidity bins while producing excesses in others. For $J/\psi$, double-resolved production dominates the additional contribution for the inputs considered, with direct and triple-resolved terms also retained. In the small-$x_g$ range examined, varying $\xgc$ mainly changes the magnitude rather than the rapidity shape, leaving limited freedom to alter the relative central and forward yields. The added terms in both processes depend on the adopted photon-fraction bound; the resolved $J/\psi$ comparison additionally depends on the small-$x_g$ extrapolation and the treatment of the experimental selections.

The leptonium rates increase as softer photons are included, with different mass and phase-space dependence for positronium, dimuonium, and tauonium. The resulting dimuonium yield remains sensitive to the low-energy input, while its electronic decay provides an experimentally useful channel. Triphoton fusion therefore merits further study as a complementary mechanism for QED bound-state production in UPCs.

\vspace{1.0cm} {\bf Acknowledgments}

This work was supported in part by the National Key Research and Development Program of China under Contracts No.~2025YFA1613900, and by the National Natural Science Foundation of China(NSFC) under the Grants 12475087 and 12235008. 

\FloatBarrier

\appendix
\section{Derivation of the Three-Body Effective Cross Section}
\label{app:wavepackets}
\setcounter{equation}{0}
\renewcommand{\theequation}{A\arabic{equation}}
\renewcommand{\theHequation}{A.\arabic{equation}}
In this appendix, we present the derivation of the effective three-body cross section starting from the $S$-matrix formalism with wave-packet initial states. The incoming state is constructed as
\begin{equation}
|\Psi\rangle_{\rm in}=\int\left[\prod_{i=1}^3
\frac{\dd^3k_i\,\phi_i(\bm k_i)}{(2\pi)^3\sqrt{2E_i}}\right]
 e^{-i\bm b_1\cdot\bm k_1}e^{-i\bm b_2\cdot\bm k_2}
|k_1k_2k_3\rangle_{\rm in},
\label{eq:a1}
\end{equation}
Summing over the final-state momenta gives the transition probability
\begin{equation}
\mathcal P(\bm b_1,\bm b_2)=
\int\left[\prod_f\frac{\dd^3p_f}{(2\pi)^3\,2E_f}\right]
\left|{}_{\rm out}\langle p_1p_2\cdots|\Psi\rangle_{\rm in}\right|^2.
\label{eq:a2}
\end{equation}
The effective cross section is defined as
\begin{equation}
\Sigma_3=\int\dd^2b_1\dd^2b_2\,
\mathcal P(\bm b_1,\bm b_2).
\label{eq:a3}
\end{equation}
Squaring the $S$-matrix element and using the normalization of asymptotic states gives
\begin{equation}
\begin{aligned}
\Sigma_3={}&\int\left[\prod_f\frac{\dd^3p_f}{(2\pi)^3 2E_f}\right]
\left[\prod_{i=1}^3
\frac{\dd^3k_i\,\phi_i(\bm k_i)}{(2\pi)^3\sqrt{2E_i}}\,
\frac{\dd^3\bar k_i\,\phi_i^*(\bar{\bm k}_i)}{(2\pi)^3\sqrt{2\bar E_i}}\right]\\[5pt]
&\times\int\dd^2b_1\,\dd^2b_2\,
 e^{-i\bm b_1\cdot(\bm k_1-\bar{\bm k}_1)}
 e^{-i\bm b_2\cdot(\bm k_2-\bar{\bm k}_2)}\\[5pt]
&\times\bigl({}_{\rm out}\langle p_1p_2\cdots|k_1k_2k_3\rangle_{\rm in}\bigr)\,
 \bigl({}_{\rm out}\langle p_1p_2\cdots|
\bar k_1\bar k_2\bar k_3\rangle_{\rm in}\bigr)^*.
\end{aligned}
\label{eq:a4}
\end{equation}
The integration over impact parameters yields
\begin{equation}
(2\pi)^4\delta^{(2)}(\bm k_{1\perp}-\bar{\bm k}_{1\perp})
\delta^{(2)}(\bm k_{2\perp}-\bar{\bm k}_{2\perp}),
\label{eq:a5}
\end{equation}
\nopagebreak[4]
which constrains the transverse momenta in the amplitude and its complex conjugate.

The energy and longitudinal-momentum constraints fix two of the remaining conjugate longitudinal integrations, leading to
\begin{equation}
\begin{aligned}
&\dd^3\bar k_1\,\dd^3\bar k_2\,\dd^3\bar k_3\,
\delta^{(2)}(\bm k_{1\perp}-\bar{\bm k}_{1\perp})\,
\delta^{(2)}(\bm k_{2\perp}-\bar{\bm k}_{2\perp})\\
&\qquad\times\bigl({}_{\rm out}\langle p_1p_2\cdots|
\bar k_1\bar k_2\bar k_3\rangle_{\rm in}\bigr)^*\\[5pt]
&=\dd^3\bar k_1\,\dd^3\bar k_2\,\dd^3\bar k_3\,
\delta^{(2)}(\bm k_{1\perp}-\bar{\bm k}_{1\perp})\,
\delta^{(2)}(\bm k_{2\perp}-\bar{\bm k}_{2\perp})\\
&\qquad\times[-i\Mcal^*(\bar k_i\to p_f)](2\pi)^4
\delta^{(4)}\!\left(\sum_{i=1}^3\bar k_i-\sum_f p_f\right)\\[5pt]
&=\dd\bar k_{1z}\,\dd\bar k_{2z}\,\dd\bar k_{3z}\,
\delta\!\left(\sum_{i=1}^3\bar k_{iz}-\sum_f p_{fz}\right)\,
 \delta\!\left(\sum_{i=1}^3\bar E_i-\sum_f E_f\right)\\
&\qquad\times(2\pi)^4[-i\Mcal^*(\bar k_i\to p_f)]\\[5pt]
&=\frac{\dd\bar k_{3z}}{|v_1-v_2|}\,(2\pi)^4
[-i\Mcal^*(\bar k_i\to p_f)].
\end{aligned}
\label{eq:a6}
\end{equation}
where $v_i=\partial E_i/\partial k_{iz}$ and the factor $|v_1-v_2|$ arises as the Jacobian of the constraint equations.
\par
\begingroup\interlinepenalty=10000\postdisplaypenalty=10000
In the narrow-wave-packet approximation, the remaining conjugate amplitude is evaluated at $\bar k_{3z}=k_{3z}$. Collecting all contributions, the differential effective cross section takes the form
\begin{equation}
\dd\Sigma_3=\frac{|\Mcal|^2}{4E_1E_2|v_1-v_2|}
\frac{\dd k_{3z}}{(2\pi)2E_3}\,\dd\Phi_n,
\label{eq:a7}
\end{equation}
\nopagebreak[4]
which is the result quoted in the main text.
\par\endgroup

\end{document}